%% file: wise_cluster.tex
\documentclass[useAMS,usenatbib]{mn2e}
\usepackage{graphicx}
\usepackage{natbib}
\usepackage{lscape}
\usepackage{longtable}
\usepackage{subfig}
\usepackage{amssymb,amsmath}
\usepackage[T1]{fontenc}
\usepackage{hyperref}
\usepackage{amsmath}
\usepackage{breakurl}
\usepackage{times}

\input{commands_mnras.tex}

\title[Clustering of Type1/2 QSOs]{The Angular Clustering of Infrared-Selected Obscured and Unobscured Quasars}
\author[DiPompeo et al.]{M.A. DiPompeo$^1$, A.D. Myers$^1$, R.C. Hickox$^2$, J.E. Geach$^3$, K.N. Hainline$^2$\\
$^1$ University of Wyoming, Dept. of Physics and Astronomy 3905, 1000 E. University, Laramie, WY 82071, USA \\
$^2$ Dartmouth College, Dept. of Physics and Astronomy, 6127 Wilder Laboratory, Hanover, NH 03755, USA  \\
$^3$ Centre for Astrophysics Research, Science \& Technology Research Institute, University of Hertfordshire, Hatfield, AL10 9AB, UK}

\begin{document}
\date{Accepted 2014 June 03; Received 2014 June 02; in original form 2014 April 09}

\pagerange{\pageref{firstpage}--\pageref{lastpage}} \pubyear{2012}

\maketitle

\label{firstpage}

\begin{abstract}
Recent studies of luminous infrared-selected active galactic nuclei (AGN) suggest that the reddest, most obscured objects display a higher angular clustering amplitude, and thus reside in higher-mass dark matter halos.  This is a direct contradiction to the prediction of the simplest unification-by-orientation models of AGN and quasars.  However, clustering measurements depend strongly on the ``mask'' that removes low-quality data and describes the sky and selection function.  We find that applying a robust, conservative mask to \textit{WISE}-selected quasars yields a weaker but still significant difference in the bias between obscured and unobscured quasars.  These findings are consistent with results from previous \textit{Spitzer} surveys, and removes any scale dependence of the bias.  For obscured quasars with $\langle z \rangle = 0.99$ we measure a bias of $b_q = 2.67 \pm 0.16$, corresponding to a halo mass of $\log (M_h / M_{\odot} h^{-1}) = 13.3 \pm 0.1$, while for unobscured sources with $\langle z \rangle = 1.04$ we find $b_q = 2.04 \pm 0.17$ with a halo mass $\log (M_h / M_{\odot} h^{-1} )= 12.8 \pm 0.1$.  This improved measurement indicates that \textit{WISE}-selected obscured quasars reside in halos only a few times more massive than the halos of their unobscured counterparts, a reduction in the factor of $\sim$10 larger halo mass as has been previously reported using \textit{WISE}-selected samples.  Additionally, an abundance matching analysis yields lifetimes for both obscured and unobscured quasar phases on the order of a few 100 Myr ($\sim$ 1\% of the Hubble time) --- however, the obscured phase lasts roughly twice as long, in tension with many model predictions.
\end{abstract}


\section{INTRODUCTION}

Studies of the optical spectra and spectral energy distributions (SEDs) of unobscured quasars\footnote[1]{Note that throughout the text we will use the terms ``AGN'' and ``quasar'' interchangeably, though the literature often divides these classes based on luminosity.} have led to many insights into the growth of supermassive black holes (BHs) through cosmic history, and shed light on the physics of BH accretion \citep{1994ApJS...95....1E, Richards:2006p3932, 2010ApJ...719.1315K, 2012NewAR..56...93A}.  Surveys in the X-ray and optical have shown that quasar activity and therefore BH growth peaks at redshift $z \sim$ 2--3 \citep{2004MNRAS.349.1397C, 2005MNRAS.360..839R, 2006AJ....131.1203F}.  Spatial clustering measurements illustrate that at all redshifts ($0 < z < 5$), quasars are found in characteristic dark matter halos of mass $\sim 3 \times 10^{12}h^{-1} \ M_{\odot}$ \citep[e.g.][]{2004MNRAS.355.1010P, 2005MNRAS.356..415C, 2007ApJ...654..115C, 2007ApJ...658...85M, 2008MNRAS.383..565D, 2009MNRAS.397.1862P, 2009ApJ...697.1634R, 2010ApJ...713..558K, 2013ApJ...778...98S}.  These results have resulted in the development of models where the processes that fuel BH growth are tied to the growth of large-scale structure in the Universe \citep{2008ApJS..175..356H, 2009MNRAS.394.1109C, 2010MNRAS.405L...1B}, and suggest that quasars play a role in regulating star formation and the emergence of the red galaxy population in halos of similar mass \citep[e.g.][]{2008ApJ...672..153C, 2008ApJ...682..937B, 2009ApJ...696..620C, 2010ApJ...719...88T, 2011MNRAS.413.2078R, 2013ApJ...778...93T, 2013MNRAS.431.3045H}.

While the previous results have provided significant advances in our knowledge of BH growth, they are only part of the story.  For a long time it has been known that a significant fraction of BH growth is obscured by gas and dust \citep[e.g.][]{1989A&A...224L..21S, 1995A&A...296....1C}.  Only recently, using techniques developed by combining \textit{Spitzer} and optical spectroscopic, X-ray, and radio surveys \citep{2004ApJS..154..166L, Stern:2005p2563} and applying them to data from the \textit{Wide-field Infrared Survey Explorer} \citep[\textit{WISE};][]{2010AJ....140.1868W}, that these obscured quasars have been found in significant numbers \citep[e.g.][]{2007ApJ...671.1365H, 2012ApJ...753...30S, 2012MNRAS.426.3271M}.  The properties of obscured quasars are being studied with increasing frequency \citep[e.g.][Hainline et al., in prep]{2011ApJ...731..117H, 2013ApJ...772...26A, Donoso:2013vz}.

Studies with \textit{Spitzer} and \textit{WISE} indicate that obscured quasars represent a large fraction of the massive BH growth in the Universe \citep[e.g.][]{Lacy:2013ws}, but their exact nature is still not clear.  Typically, the obscuring material is attributed to either a ``dusty torus'' \citep[an axisymmetric structure intrinsic to the nuclear region, e.g.][]{1993ARA&A..31..473A}, or larger scale, high covering fraction obscuring material \citep[e.g.][]{2004ApJ...611L..85P, 2012ApJ...755....5G}.  The former fits into the simplest unified model for AGN, in which the difference between obscured and unobscured sources is due simply to source orientation --- at small viewing angles to the symmetry axis, one has a clear view to the nucleus, while at larger viewing angles the torus blocks our view.  While observations support this model at low-$z$ and low-luminosity \citep[particularly in Seyfert galaxies;][]{1993ARA&A..31..473A}, it is unclear whether this model applies to objects with quasar luminosities.  In contrast, high-covering fraction explanations could be due to quasar fueling by major mergers of galaxies, which drive gas and dust clouds to the nucleus, obscuring AGN activity.  This hypothesis is suggested by models of BH-galaxy co-evolution \citep[e.g.][]{1988ApJ...325...74S, 2005Natur.433..604D, 2008ApJS..175..356H}.

A powerful way to test these scenarios is by examining the environments of quasars, specifically the masses of their parent dark matter halos.  The simple unified model robustly predicts no difference between the environments of obscured and unobscured quasars.  A difference in halo mass \textit{is} expected in some evolutionary, merger-driven scenarios.  If obscured quasars are an early growth phase then they are in the process of ``catching up'' to their final mass relative to their host galaxy and halo \citep[e.g.][]{2010MNRAS.408L..95K}, and obscured quasars would occupy larger mass halos compared to unobscured quasars of the same luminosity and BH mass.  

A common method to estimate the typical halo mass of a population of quasars is to measure their spatial clustering.  In current cosmological models the Universe is dominated by dark matter, with galaxies embedded in halos that contain the vast majority of the mass that drives their clustering.  Populations of galaxies can be related to underlying halos that have some characteristic mass using models of how dark matter halos collapse at different mass thresholds \cite[e.g.][]{2001MNRAS.323....1S, 2005ApJ...631...41T, 2010ApJ...719...88T}.  Since most, if not all, galaxies contain a supermassive BH whose properties correlate with properties of the host galaxy \citep[e.g.][]{1998AJ....115.2285M, 2000ApJ...539L...9F, 2000ApJ...539L..13G, 2002ApJ...574..740T, 2007ApJ...663...61L, 2012MNRAS.419.2497B, 2013ApJ...773....3C, 2014ApJ...782....9H}, quasars are thought to be a phase in the lifetime of all galaxies.  By measuring quasar bias ($b_q$), or how strongly quasars cluster relative to an underlying model of the clustering of dark matter (for a given cosmology), it is possible to measure characteristic halo masses.  With samples derived from \textit{Spitzer} and \textit{WISE}, these measurements have recently been made for the first time for obscured quasars.

\cite{2011ApJ...731..117H} made the first comparison of the clustering of IR-selected obscured and unobscured quasars using a \textit{Spitzer}-selected sample in the 9 deg$^2$ Bo\"{o}tes field \citep{2007ApJ...671.1365H}.  Employing the technique of \cite{2009MNRAS.399.2279M}, which uses the full probability distribution function of photometric redshifts to calculate 3D clustering, evidence was found that obscured quasars may cluster more strongly than their unobscured counterparts, and thus reside in higher mass halos.  This raised the possibility that obscured quasars are indeed an early evolutionary phase of black hole growth \citep[as in Figure 1 of][]{2008ApJS..175..356H}, though the difference in clustering magnitude was only marginally significant.  \citet[hereafter D13]{Donoso:2013vz} performed a similar measurement for a much larger \textit{WISE}-selected sample, covering an area of $\sim$3600 deg$^2$ overlapping the Sloan Digital Sky Survey \citep[SDSS;][]{2000AJ....120.1579Y} footprint.  They find a far larger and more significant difference in bias (and thus halo mass) than \cite{2011ApJ...731..117H}.

These findings have important implications and need to be explored further, in large part because clustering measurements can suffer from systematic effects.  Even in regions where the data are relatively uniform in terms of depth, such as the region chosen by D13, clustering results remain highly dependent on the details of the sample ``mask'', or the areas on the sky that remain after removing regions of bad or unusable data.  As we will demonstrate here, small changes in the mask \citep[or even in the weighting of the mask, e.g.][]{2011MNRAS.417.1350R, 2013MNRAS.435.1857L} have the potential to lead to large changes in the clustering amplitude.  Additionally, under the assumption that quasars cluster like dark matter, the quasar bias is roughly scale-independent, at least on large scales.  This has been seen empirically in many studies --- however, there appears to be a somewhat strong scale dependence in the results of D13 (especially for obscured quasars), which may indicate insufficient masking of the data.

Here we present an independent analysis of the obscured and unobscured \textit{WISE}-selected quasar angular clustering, in the same region as D13.  We build our own mask for the data, paying particular attention to the effects of differences in the mask on the final clustering measurement.  We use a cosmology where $H_0 = 71$ km s$^{-1}$ Mpc$^{-1}$, $\Omega_{\textrm{M}}=0.27$, $\Omega_{\Lambda} =0.73$, $\Omega_{\textrm{b}} = 0.045$ and $\sigma_8 =0.8$ for all calculated parameters \citep{2011ApJS..192...18K}.  All magnitudes are given in the Vega system.

\section{DATA}

\subsection{\textit{WISE}}
Our sample is selected from the all-sky catalog of \textit{WISE}.  \textit{WISE} mapped the sky multiple times (on average $\sim$10 times in regions away from the ecliptic poles, where coverage increases due to the observing strategy) in four bands at 3.4, 4.6, 12 and 22 $\mu$m, which are referred to as $W1$, $W2$, $W3$ and $W4$, respectively.  The 5$\sigma$ sensitivity limit in each band is at least 0.08, 0.11, 1 and 6 mJy, respectively, and improves in areas of higher coverage.  The angular resolutions are 6.1$''$, 6.4$''$, 6.5$''$, and 12$''$, respectively.  An object is included in the all-sky catalog if it is detected at SNR$>$5 in at least one band, has at least five good measurements, and is not flagged as a spurious source in at least one band (see the \textit{WISE} All-Sky Release Explanatory Supplement\footnote[2]{\url{http://irsa.ipac.caltech.edu/Missions/wise.html}}).

The mid-IR wavelengths and large area of \textit{WISE} are ideal for uniformly selecting both obscured and unobscured AGN in large numbers.  AGN are redder than normal galaxies at these wavelengths, because the black-body spectrum of stellar populations peak at near-IR wavelengths ($\sim1.5 \mu$m) while the hot dust in AGN causes a rising power-law spectrum at longer wavelengths.  This was first illustrated with \textit{Spitzer} \citep[e.g.][]{2004ApJS..154..166L, Stern:2005p2563, 2007ApJ...660..167D}, and more recently with $W1$ and $W2$ from \textit{WISE} \citep{2012ApJ...753...30S, 2012MNRAS.426.3271M, 2013MNRAS.434..941M, 2013ApJ...772...26A}.  The most significant contaminants in this selection are cool brown dwarfs \citep{2011ApJS..197...19K} and high redshift galaxies.  The latter can be largely eliminated by making a flux or magnitude cut, which we apply as described below.

\citet{2012ApJ...753...30S} showed that a simple color cut at $W1 - W2 > 0.8$ for objects with $W2 < 15.05$ (the 10$\sigma$ flux limit in this band) identifies AGN at 80\% completeness and a contamination rate of 5\% 
\citep[when compared with \textit{Spitzer} selection;][]{Stern:2005p2563}.  As in D13, we apply these criteria to the \textit{WISE} all-sky data (in addition to a cut at $W2 > 10$, which removes 994 objects, only 84 of which would make it through the rejection described in the next section) in the region $135^{\circ} < RA < 226^{\circ}$ and $1^{\circ} < DEC < 54^{\circ}$ (see the next section), and identify 249,169 AGN candidates.  Note that we do not correct the \textit{WISE} photometry for Galactic extinction.  Extrapolating the plummeting near-IR extinction curves of \citet{2009ApJ...699.1209F} suggests that extinction coefficients are $< 0.2$ in $W1$ and $W2$, and $E(B-V)$ is low in this region (and regions of high extinction are eliminated from the sample; see \S2.2).

\subsection{Data Rejection and the Angular Mask}
After selecting all potential \textit{WISE} AGN, we limit our study to the same region as D13, chosen because it is relatively free of contamination from the Galactic plane, is sufficiently far from the ecliptic pole, and it overlaps with SDSS imaging.  This region is between $135^{\circ} < \textrm{RA} < 226^{\circ}$ and $1^{\circ} < \textrm{DEC} < 54^{\circ}$, for a total of 4,127 deg$^{2}$.  However, not all of this region is free from contamination, and so we build an independent angular mask using the spherical cap utility \textsc{mangle}\footnote[3]{\url{http://space.mit.edu/\~molly/mangle/}} \citep{2004MNRAS.349..115H, 2008MNRAS.387.1391S} to remove regions of bad/compromised data.  Below we detail the exact components that go into making the mask and cleaning the sample\footnote[4]{\textsc{mangle} polygon files marking the regions of data that have been removed can be found at \url{http://faraday.uwyo.edu/~admyers/wisemask2014/wisemask.html}}.

\begin{figure}
   \includegraphics[width=8cm]{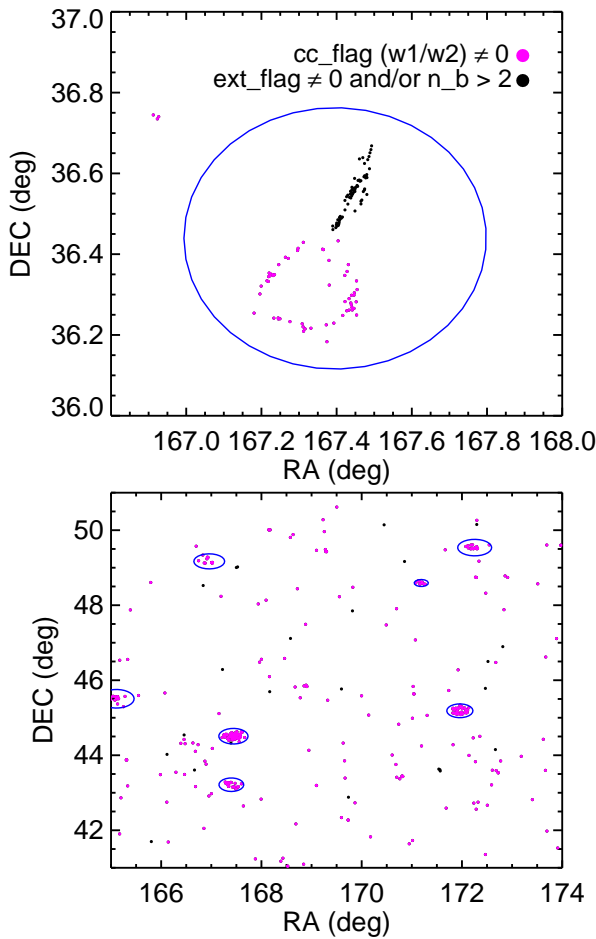}
  \caption{Both panels show objects flagged in \textit{WISE}, and the results of our algorithm to identify tightly grouped flagged objects that need the full area (solid blue ellipses) around them removed, and not just the points themselves. Magenta points are those flagged with \textsc{cc\_flags} $\neq$ 0 in $W1$ or $W2$, black points are those with \textsc{ext\_flag} $> 1$ and/or deblending flag \textsc{n\_b} $> 2$.  Magenta points are automatically removed from the final sample, black points only if they fall within the blue circles that are part of our final mask.\textit{Top:} Close-up of a region with multiple flag types. \textit{Bottom:} A 10-by-10 degree region showing what kinds of regions our algorithm identifies and removes.  It may be conservative in a few cases, such as the region at the top left, but overall does a fair job of identifying highly clustered, flagged objects.}
  \label{fig:masks}
\end{figure}

\begin{itemize}
\item Regions of high Galactic extinction are excluded, as these can impact clustering measurements, particularly for faint objects \citep{2006ApJ...638..622M, 2011MNRAS.417.1350R}.  We build a grid of points spaced by 0.5$^{\circ}$ in RA and DEC, and use the dust maps of \citet{1998ApJ...500..525S} to find points where $A_g > 0.18$ \citep{2006ApJ...638..622M}.  We remove circular regions around these points with radii of 0.36$^{\circ}$ from the final mask.

\item \textit{WISE} tiles with significant contamination from the Moon (\textsc{moon\_lev} $> 1$ in $W4$ in the \textit{WISE} flags) are excluded.  Each \textit{WISE} tile is 1.56$^{\circ}$ on a side.  For simplicity in converting our mask between coordinate systems, we remove circular regions, even though the tiles are rectangular.  The regions are centered on the position of the tile with a radius of 1.1$^{\circ}$ (the length of half of the diagonal of a tile).

\item As described in the next two sections, imaging from SDSS is used to split the \textit{WISE}-selected AGN into obscured and unobscured samples.  Therefore for the final sample, bad fields in the SDSS data are removed, and the SDSS bright star mask is applied \citep[e.g.][]{2011ApJ...728..126W, 2012MNRAS.424..933W}\footnote[5]{see also the SDSS-III Data Model, \burl{http://data.sdss3.org/datamodel/files/BOSS_LSS_REDUX/reject_mask/MASK.html}}.

\item After removing the above regions from the mask, there are still clearly quite a few artifacts remaining in visual inspections of the data, particularly highly clustered objects that are likely galaxies and other resolved objects broken up into point sources by \textit{WISE}.  In order to remove these, we developed a method to locate highly grouped objects \textit{in the flagged data} via a friend-of-friends type algorithm (Figure~\ref{fig:masks}).  We use objects that are flagged as having compromised photometry due to diffraction spikes, optical ghosts, persistence, or scattered light (\textsc{cc\_flags} $\neq 0$ in $W1$ or $W2$), as extended (\textsc{ext\_flag} $\neq 0$), or significantly deblended (\textsc{n\_b} $ > 2$).  Figure~\ref{fig:masks} illustrates that the algorithm is tuned conservatively such that it may mask more regions than necessary, but overall it does an excellent job of removing the full area around contamination.  We note that simply removing these flagged data is not sufficient --- both because without fully incorporating them into the mask, the resulting random catalog does not accurately mimic the data in these regions, and because objects in the area but not necessarily flagged by \textit{WISE} may still be affected.  This procedure removes the regions around objects resolved by \textit{2MASS} that D13 remove from their mask, in addition to others.  We analyze our results both with and without this component of the mask applied to illustrate the effects including these regions --- we refer to the sample including this mask as using the ``full mask'', and  as the ``partial mask'' when we do not remove these regions.

\item Even after removing the regions in the point above, the 7617 remaining objects with \textsc{cc\_flags} $\neq 0$ are \textit{not} randomly distributed on the sky, though a visual inspection suggests that they may be.  A clustering analysis on these flagged points verifies that they are not random.  Therefore, for completeness we also incorporate into our mask small circular regions with radii of 1$'$ around each of these points.  We note that simply removing these points or including this small addition to the mask makes no difference in the results (unlike for the highly clustered points in the point above).  In the case of results using the ``partial mask'', objects with \textsc{cc\_flags} $\neq 0$ are still removed from the sample, but this component of the mask is also not used.

\end{itemize}

Applying the full mask and removing the flagged data leaves us with a final sample of 177,709 \textit{WISE}-selected AGN, over an area of 3,289 deg$^{2}$.

\subsection{SDSS}
The SDSS has imaged roughly a quarter of the sky in five optical bands ($ugriz$).  We will utilize the $r$-band in this work (see below), which reaches 50\% completeness at $r=22.6$ \citep{2009ApJS..182..543A}.  We adopt the SDSS pipeline $psfMag$ values, as we are interested in isolating the AGN contribution to the flux as much as possible, which is unresolved, while a significant fraction of source host galaxies are resolved in the SDSS imaging.  This is in contrast to D13, who use $modelMags$ from SDSS.  However, as a test we have performed our analysis using the pipeline $modelMags$ as well, and find that the results are not substantially different.  We correct magnitudes for Galactic extinction using the values supplied with the SDSS data \citep{1998ApJ...500..525S}.  SDSS magnitudes are converted from the (nearly) AB magnitude system to the Vega system, as the \textit{WISE} magnitudes are supplied in Vega mags.  We use a simple conversion factor, $m_{r, AB} = m_{\textrm{r, Vega}} + 0.16$ \citep{2007AJ....133..734B}.

\subsection{Obscured and Unobscured AGN}
In a detailed,  multi-wavelength study of the Bo\"{o}tes field, \citet{2007ApJ...671.1365H} found that an optical/IR color cut at $R- [4.5] = 6.1$ (Vega) robustly separates the obscured and unobscured AGN populations.  As $W2$ closely resembles the \textit{Spitzer} 4.6$\mu$m band, this can be directly applied to objects with both SDSS $r$-band and $W2$ data, as we have here.

To make this color separation, we match the 177,709 \textit{WISE}-selected AGN from above to the SDSS DR8 catalog, using a 2$''$ radius, accepting only the closest match.  We find 147,251 (83\%) matches, leaving 30,458 (17\%) with no SDSS counterparts.  The sources with no matches are randomly distributed on the sky, and we have no reason to believe that any significant portion of them are undetected in SDSS because they are artifacts in \textit{WISE}.  D13 performed a visual inspection of many of these objects in the COSMOS field, which indicate that they are bona-fide AGN with X-ray counterparts in many cases.  As a check, we also perform our analysis only including sources with SDSS counterparts and find that the results remain the same, as discussed in section 4.

We relax the \citet{2007ApJ...671.1365H} color cut to $r-W2 > 6$ to split our sample into obscured  and unobscured subsamples (and also remove 80 sources for which $r_{psf} < 15$ or $r_{psf} > 25$).  All sources without an SDSS match are placed into the obscured sample, resulting in sample sizes of 74,889 (42\%) and 102,740 (58\%) for ``obscured'' and ``unobscured'', respectively.  This unobscured fraction is slightly higher than what is found in D13; this is likely because the additional removal of regions around objects flagged by \textit{WISE} tends to remove more ``obscured'' sources.  As we will see, this effect has significant consequences for the final angular clustering measurements.

\begin{figure}
   \includegraphics[width=9cm]{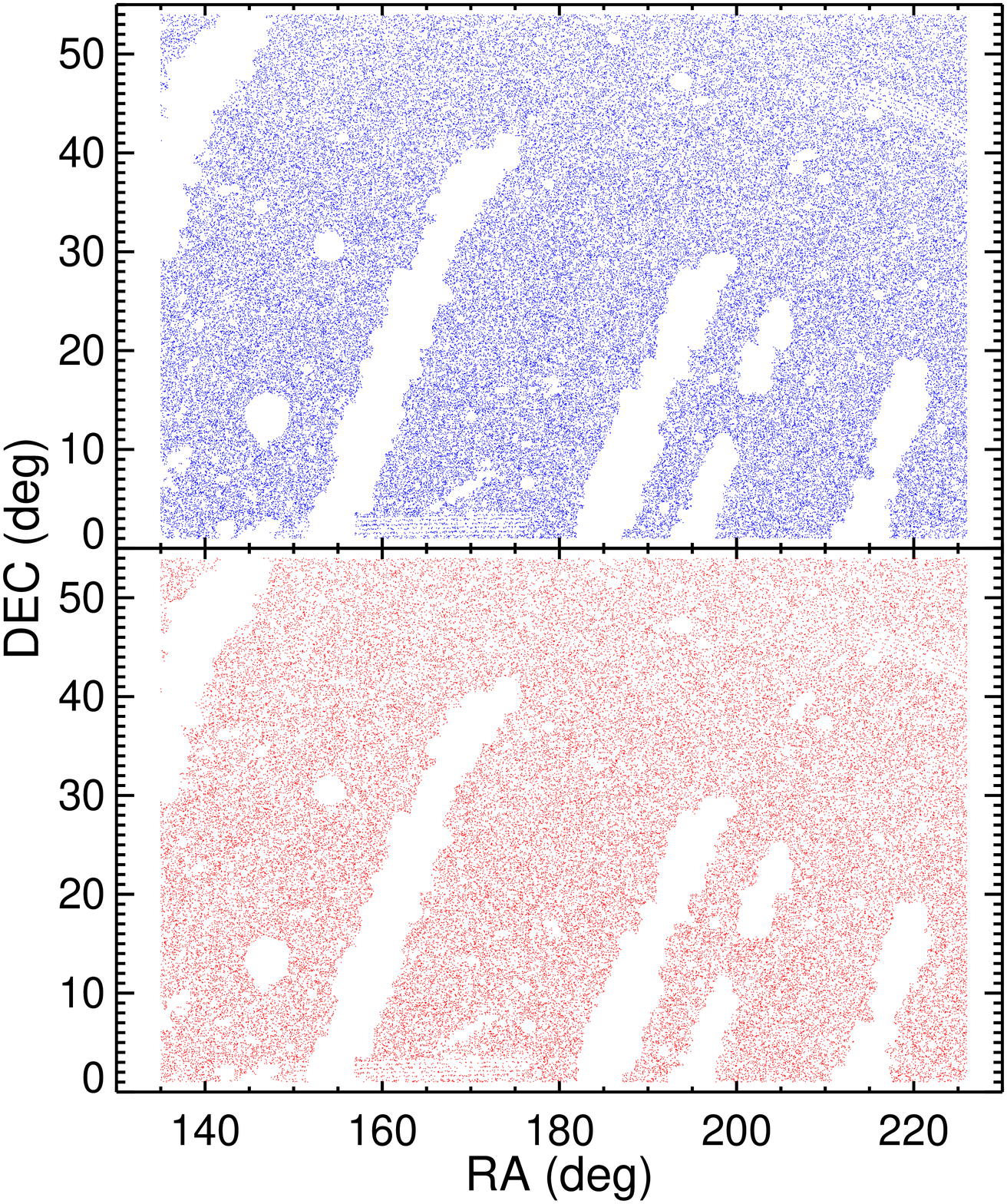}
  \caption{The \textit{WISE}-selected AGN, with the mask applied (see \S2.2), and split by unobscured (top, blue; 102,740 objects) and obscured (bottom, red; 74,889 objects) subsamples (see \S2.4).
 \label{fig:samples}}
\end{figure}

\begin{figure}
   \includegraphics[width=8cm]{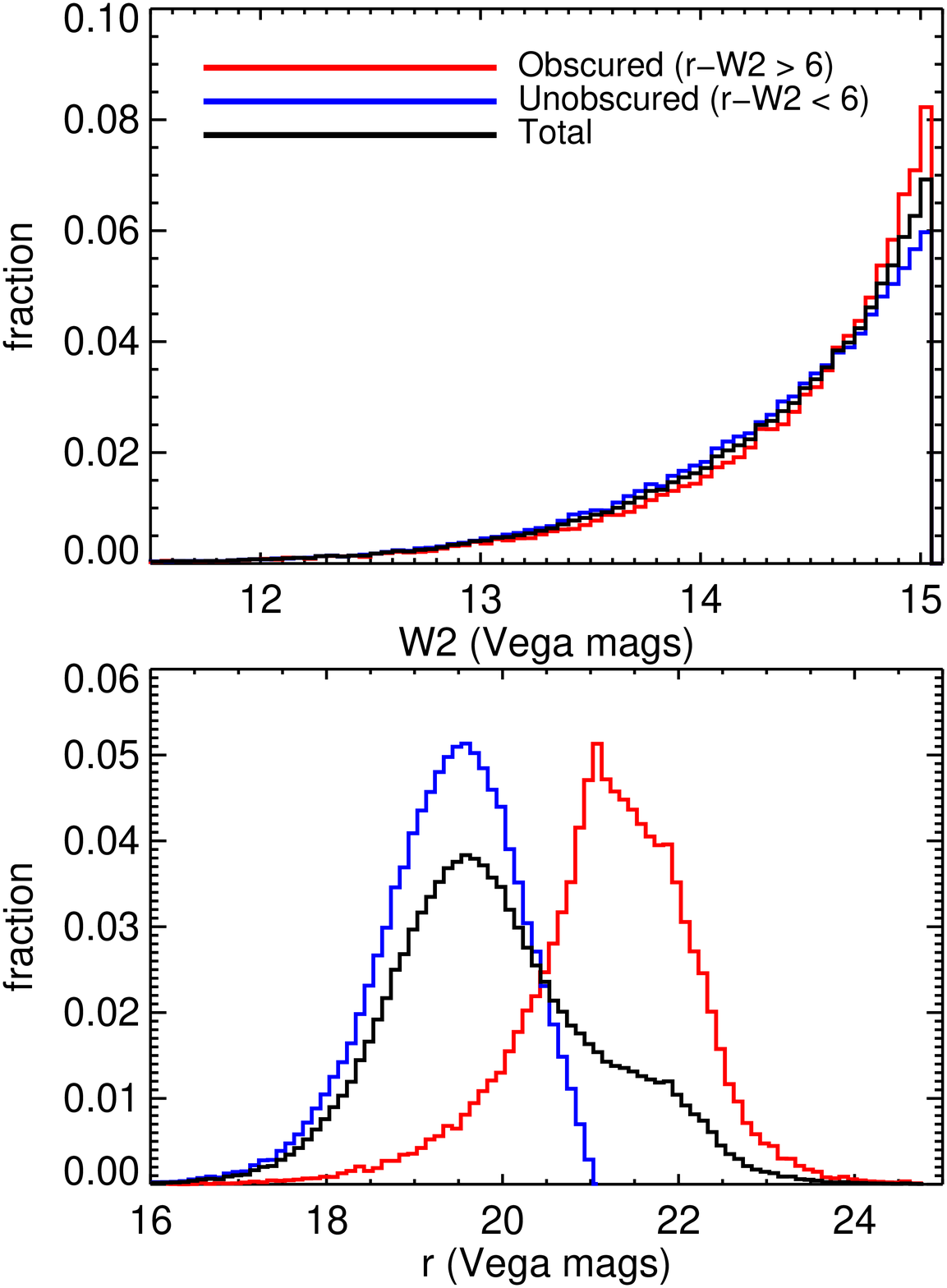}
  \caption{\textit{Top:} The $W2$ magnitude distributions for the obscured (red), unobscured (blue) and total (black) samples.  \textit{Bottom:} The SDSS $r$-band magnitudes, corrected for Galactic extinction and converted to the Vega system, for the total sample and split by obscured/unobscured AGN.\label{fig:photometry}}
\end{figure}

\begin{figure}
   \includegraphics[width=8cm]{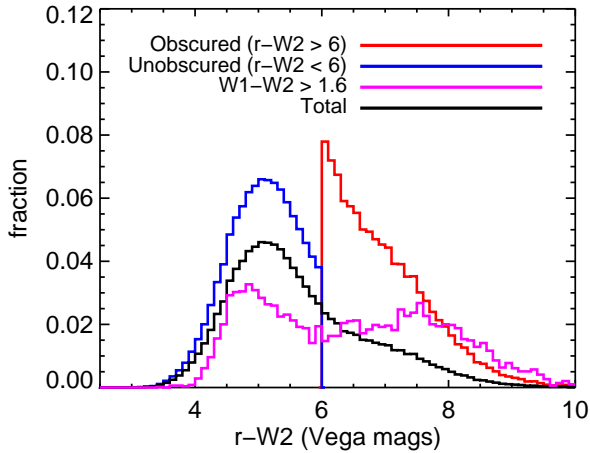}
  \caption{The optical-IR ($r - W2$) color distributions of the full sample of \textit{WISE}-selected AGN with SDSS counterparts (black), the obscured sample (red) and the unobscured sample (blue).  By definition, the obscured/unobscured split is at $r-W2=6$, which is very close to what was found by Hickox et al.\ (2007) and clearly seen as the location of the bi-modality in color when the reddest sources in \textit{WISE} are considered ($W1-W2 > 1.6$).\label{fig:colors}}
\end{figure}

The distributions on the sky of the two final samples are shown in Figure~\ref{fig:samples}. The $r$ and $W2$ distributions of these samples are shown in Figure~\ref{fig:photometry}.  While the obscured objects are slightly fainter in $W2$ on average, the shapes of the $W2$ distributions are very similar between the samples suggesting that the samples are not strongly biased by the $r-W2$ color cut.  The main difference comes from the $r$ magnitudes, where the obscured objects (with SDSS matches --- those with no match are not shown in this figure) are nearly 2 magnitudes fainter, on average.  Figure~\ref{fig:colors} shows the $r-W2$ color distribution for the samples.  To illustrate the validity of the optical/IR color cut, we show in this figure the color distribution for the reddest sources in \textit{WISE} ($W1-W2 > 1.6$) --- it is clearly bi-modal with a minima at $r-W2 \approx 6$.  We note that for heavily obscured quasars, even the PSF optical magnitudes may be dominated by starlight, meaning that the $r-W2$ color represents only a lower limit on the color of the AGN component.  These magnitude and color distributions are nearly identical to those in D13 --- despite our differences in building the mask, the overall photometric properties of the samples are the same.

\subsection{Morphology}
As noted by D13, the optical morphology of IR-selected AGN can provide insight into the general quasar population.  We refer the reader to D13 for a robust analysis of the morphologies of \textit{WISE}-selected AGN using \textit{HST-COSMOS} data, and perform a simple check of their results here.  If we simply use the SDSS pipeline \textsc{obj\_type} keyword \citep[see Table 6 of][]{2002AJ....123..485S}, we find that of the IR-selected AGN in our sample that are detected in SDSS, 65\% are unresolved.  This is higher than the 55\% of D13, partly because of the higher unobscured/obscured ratio we have and the fact that unobscured sources are more likely to be unresolved (because the light from the nucleus is more likely to outshine the host galaxy).  

In addition, we also utilize the deeper imaging of SDSS Stripe 82 to analyze the morphology of IR-selected AGN.  Stripe 82 does not overlap our sample at all, but we can apply the same selection criteria and masking procedure to \textit{WISE} sources in the Stripe 82 region for this analysis.  Doing so and matching to the Stripe 82 data (with a radius of 2$''$) we find 6,118 objects.  In this case, 61\% of the sources are unresolved --- as expected, more objects are resolved in this deeper data.  Applying the same optical-IR color cuts to divide these objects into obscured and unobscured subsamples, we find that 2,027 (33\%) are obscured and 4,089 (67\%) are unobscured (note that this obscured fraction does not included objects undetected in SDSS, and is therefore lower than the sample we use for our clustering measurements).  Of the obscured objects, 34\% are unresolved, while 74\% of unobscured sources are unresolved.  This latter figure is slightly higher than that of D13, but not overwhelmingly so.  Our sample follows a very similar morphological distribution as D13.  It also matches well the results of \citet{2007ApJ...671.1365H}, though their deeper optical imaging leads to a higher resolved fraction for obscured quasars.

\subsection{Redshift Distributions}
In order for an accurate comparison between obscured and unobscured sources, we must compare objects over a similar range in redshift, with a similar $dN/dz$.  To verify this we apply the same selection criteria and mask described above to the 9 deg$^{2}$ Bo\"{o}tes survey field.  There is extensive spectroscopy of AGN in this field \citep[the AGN and Galaxy Evolution Survey; AGES,][]{2012ApJS..200....8K}, as well as photometric redshift information from \textit{Spitzer} IRAC data \citep{2006ApJ...651..791B, 2011ApJ...731..117H}.  

\begin{figure}
\centering
\hspace{0cm}
   \includegraphics[width=8cm]{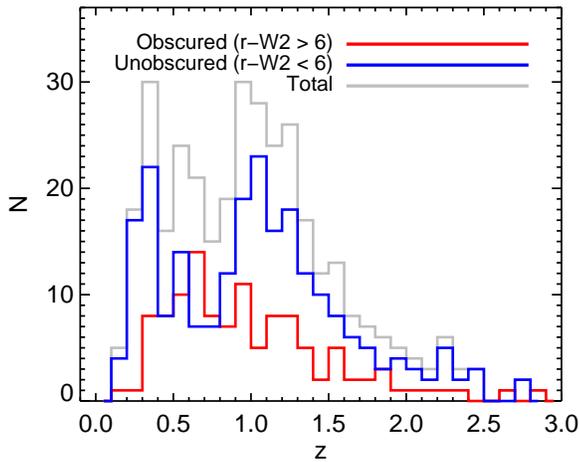}
   \vspace{0cm}
  \caption{The redshift distribution of the total, obscured and unobscured quasars, using the same selection and mask for objects in the Bo\"otes field.  Statistically, the obscured and unobscured $dN/dz$ is the same, allowing for accurate comparison of the two samples.\label{fig:z}}
\end{figure}

We find 361 \textit{WISE}-selected AGN with redshift information in this field --- 232 unobscured and 129 obscured.  One obscured source is a strong outlier from the main distribution, and has a poorly constrained redshift based on visual inspection and comparison with other photometric estimators, and so this object is removed from the analysis.  All of the unobscured sources have reliable spectroscopic  redshifts, with the exception of one which has an accurate photometric $z$.  There are 43 obscured sources with photometric redshifts, and 85 with spectroscopic.  Figure~\ref{fig:z} shows the redshift distributions of the total, obscured and unobscured AGN.  The mean/median/standard deviations of each distribution are 1.02/0.98/0.56 (total), 0.99/0.91/0.53 (obscured), and 1.04/1.04/0.58 (unobscured).  These values are in reasonable agreement, and a simple Kolmogorov-Smirnov (KS) test indicates that the obscured and unobscured samples are drawn from the same parent population.

It is possible that the obscured sources not detected by SDSS are optically faint because they are at higher redshift than the rest of the sample, and thus including these sources in our analysis could bias the redshift distribution.  As a test, we examine the redshift distribution of the objects satisfying our selection in the Bo\"{o}tes field that have SDSS $r$ magnitudes fainter than the completeness limit of SDSS ($r=22.6$; note however that there are objects in the SDSS catalog fainter than this limit).  There are 35 of these sources, and their $z$ distribution is very similar to that of the overall obscured sample with a mean/median/standard deviation $\sim$ 1.11/1.04/0.56.  We therefore find little evidence that including the objects dropping out of SDSS significantly shifts our expected $dN/dz$ to higher values.

We note that while the spike at low redshift ($z \sim 0.25$) in the unobscured sample has been seen before in samples of IR-selected quasars \citep[e.g.][]{2013ApJ...772...26A}, it raises potential concerns for the analysis.  It is possible that these objects are in fact obscured AGN that appear to be unobscured due to optical light from the host galaxy contaminating the nuclear region.  We analyze the effects of this possibility in \S4.

\section{MEASUREMENTS \& RESULTS}
\subsection{Angular Clustering}
The two-point angular correlation function ($\omega (\theta)$) is the probability that a given pair of objects with mean number density $n$, separated by a projected angular distance $\theta$, are within a solid angle $d \Omega$ \citep{1969PASJ...21..221T, Peebles:1980vn}:
\begin{equation}
dP = n(1+\omega(\theta))d\Omega.
\end{equation}
This is generally estimated by comparing the number counts of objects in annuli of increasing radii with what is expected for a completely random distribution.  For direct comparison with the results of D13, we also use the \citet{1993ApJ...412...64L} estimator:
\begin{equation}
\omega(\theta) = \frac{DD - 2DR +RR}{RR}.
\end{equation}
Here $DD$, $DR$, and $RR$ are the number of data-data, data-random, and random-random sample pairs in each bin of $\theta$ (normalized by the numbers of objects in the samples).  The random objects \textit{must} follow the same angular selection function as the data, i.e.\ they must be constrained by the same mask.  This is simple in our case as we assume that the sources are distributed evenly across the region of interest, with holes described by the \textsc{mangle} polygons discussed in \S2.2.  We use the \textsc{mangle} utility \textsc{ransack} to generate a random distribution of simulated sources that follows the same mask as the data.  The random catalog is the same for both obscured and unobscured samples (since the same mask is appled to both), and contains 1.5 million sources (more than a factor of ten larger than the data sets), to ensure that the random catalog number counts do not limit the statistical precision.

\begin{figure*}
\centering
   \includegraphics[width=16cm]{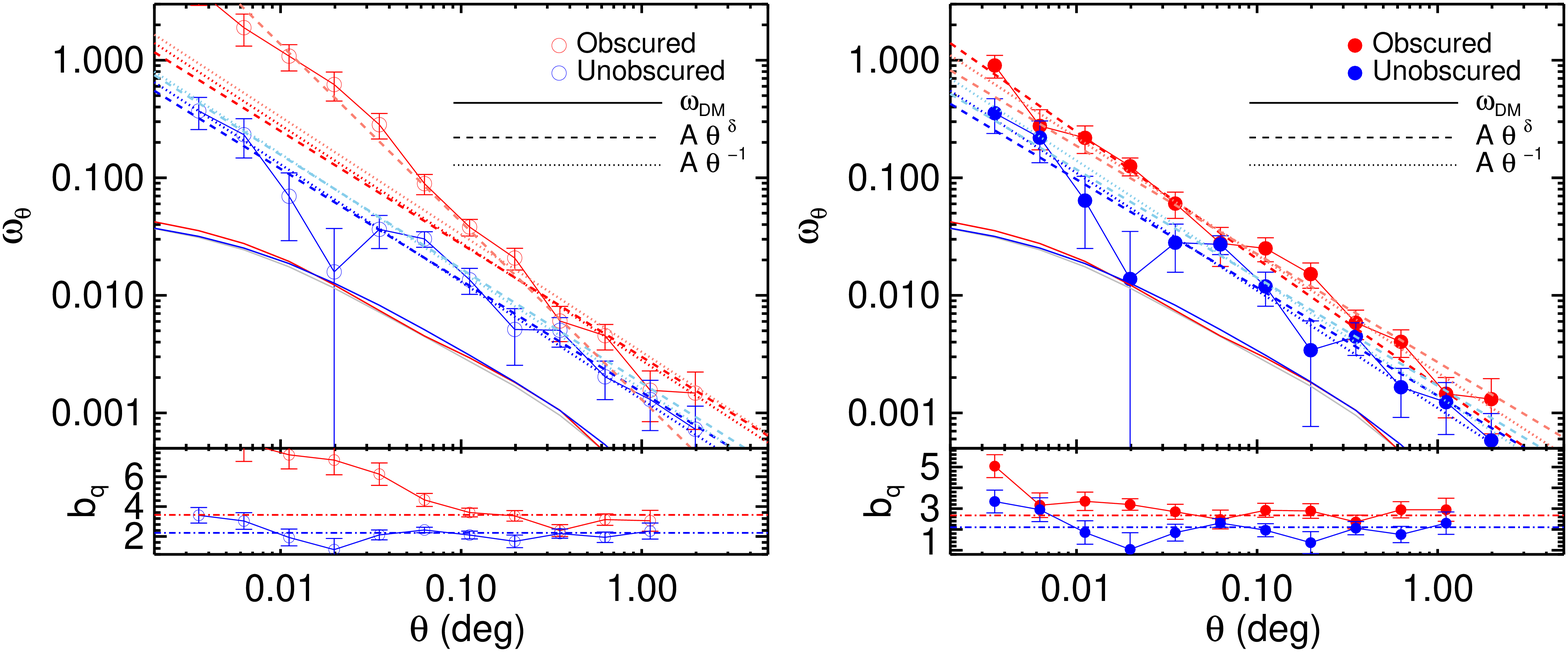}
  \caption{Angular clustering of obscured (red) and unobscured (blue) AGN, with the full mask applied (right) and without the regions around clustered, \textit{WISE} flagged data removed (left; the partial mask).  Fits are shown both for  a power-law with the slope as a free parameter (dashed lines) and with the slope fixed at $\delta = -1$ (dotted lines), as well as over the full range of data shown (darker lines) and over the range 0.2-0.5 degrees (lighter lines).  The  quasar bias ($b_q$, see \S3.3) is shown underneath each autocorrelation.  We see that applying the full mask reduces the difference in clustering amplitude and bias (and thus the inferred halo masses) between obscured and unobscured samples, as well as removes the scale dependence of the bias for obscured quasars, which is expected to be nearly flat.\label{fig:w4dex}}
\end{figure*}

We calculate the angular autocorrelations with several different binnings, and generally present those with four bins per decade (written as 4/dex in tables and figures), beginning at $\sim$0.003$^{\circ}$ ($\sim$12$''$) and extending to $\sim$2.1$^{\circ}$.  This binning provides errors at a level that allows us to perform fits to all subsamples using the full covariance matrices.  These results are shown in Figure~\ref{fig:w4dex} --- the left panel shows the angular autocorrelation using the partial mask, and the right shows the results using the full mask.

\begin{figure}
 \centering
   \includegraphics[width=7.5cm]{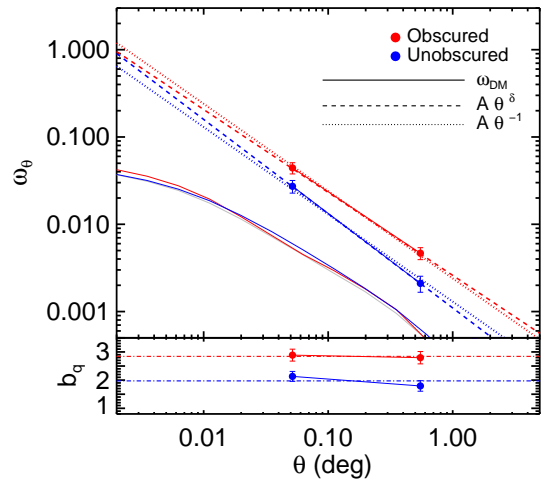}
  \caption{Angular clustering of obscured (red) and unobscured (blue) AGN with two large bins, to reduce the scatter and highlight the difference in clustering amplitude that remains even when using the full mask.  Fits are shown for both a power-law with the slope as a free parameter (dashed lines) and with the slope fixed at $\delta = -1$ (dotted lines).  The quasar bias $b_q$ is shown in the panel underneath.\label{fig:wbig}}
\end{figure}

In order to reduce the errors as much as possible and highlight any potential differences in clustering amplitude between obscured and unobscured AGN, we also calculate the angular autocorrelation using two large bins --- one extending from 0.003$^{\circ}$ to 0.1$^{\circ}$ (centered at 0.05$^{\circ}$), and the other from 0.1$^{\circ}$ to 1$^{\circ}$ (centered at 0.55$^{\circ}$).  These are plotted in Figure~\ref{fig:wbig}.

\subsection{Error Estimates and Fits}
To estimate errors on the angular clustering, we use inverse-variance-weighted ``jackknife'' resampling \citep[e.g.][]{2002ApJ...579...48S, 2005MNRAS.359..741M, 2007ApJ...658...85M}.  This method divides the data into $N$ regions, builds $N$ subsamples by iteratively \textit{removing} each region, and then repeats the clustering measurement with each subsample.  We calculate errors (and fits, see below) from the full covariance matrix generated from these jackknife iterations, using $N=16$ equal-area regions.  If each subsample is denoted by $L$, then the inverse-variance-weighted covariance matrix ($C_{ij} = C(\theta_i,\theta_j)$; $i$ and $j$ denote angular size bins) is:
\begin{multline}
C_{ij} = \sum_{L=1}^{N} \sqrt{\frac{RR_L(\theta_i)}{RR(\theta_i)}} [\omega_L(\theta_i) - \omega(\theta_i)] ~\times  \\
      \sqrt{\frac{RR_L(\theta_j)}{RR(\theta_j)}} [\omega_L(\theta_j) - \omega(\theta_j)],
\end{multline}
where $\omega$ is the angular correlation for all of the data and $\omega_L$ is the same for subsample $L$.  The $RR$ terms are the \textit{un-normalized} random-random counts, and account for the different number of objects expected in each region.  The jackknife errors ($\sigma_i$) are taken from the diagonal elements of the covariance matrix, i.e.\ $\sigma_i^2 = C_{ii}$.  These are the error bars shown in Figures~\ref{fig:w4dex} and \ref{fig:wbig}.

\begin{figure}
\centering
\hspace{0cm}
   \includegraphics[width=8cm]{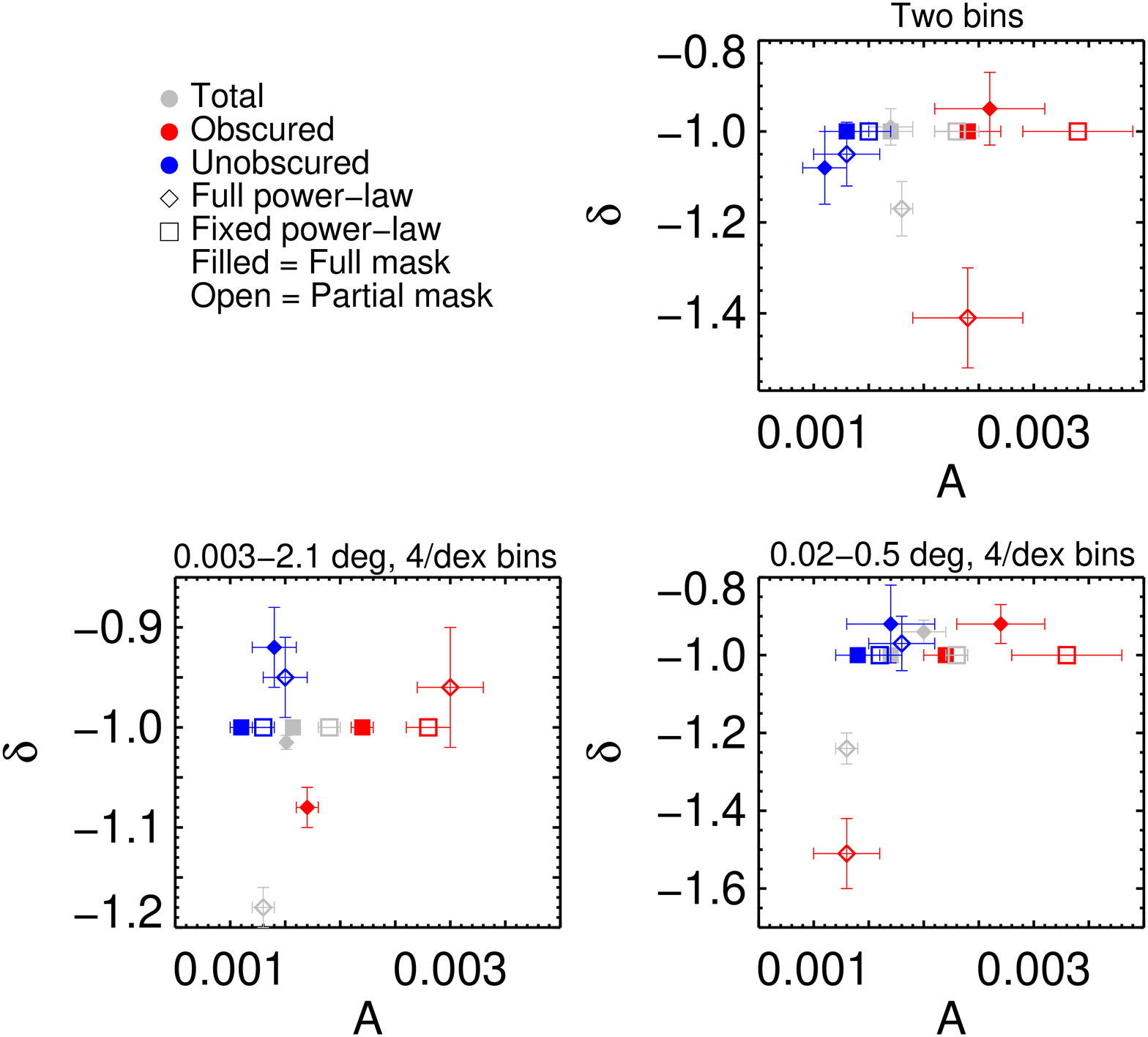}
   \vspace{0cm}
  \caption{Comparison of the angular autocorrelation power-law best-fit parameters for the total, obscured and unobscured samples.  Open symbols are for the sample with the partial mask applied, where the regions surrounding clustered \textit{WISE} flagged data are not removed, and filled symbols are fits when the full mask is applied.  The top right is when binning with two large bins, the bottom left is with a binning of four bins per dex and fit over the full range, and the bottom right is the same binning but only fit on intermediate scales.  In the case of the fixed slope power-laws (squares; $\delta = -1$), we can see that using the full mask always reduces the amplitude, especially for the obscured sources.  When leaving the power as a free parameter (diamonds), in many cases the autocorrelation of the obscured sources becomes much steeper and actually causes an increase in the clustering amplitude.  When the full mask is used, the slopes of all of the samples are much more consistent.\label{fig:fitcompare}}
\end{figure}

We perform two power-law fits to the data, of the form $\omega_m (\theta) = A \theta^{\delta}$.  In one fit, both $A$ and $\delta$ are free parameters; in the other the power is fixed at $\delta = -1$, which is a typical empirically determined value for the quasar population \citep{2006ApJ...638..622M, 2007AJ....133.2222S, 2009ApJ...697.1656S, 2009ApJ...697.1634R, 2012MNRAS.424..933W}.  Fits are performed using the full covariance matrix, using a $\chi^2$ minimization:
\begin{equation}
\chi^{2} = \sum_{i,j} [\omega (\theta_i) - \omega_m(\theta_i)] C_{ij}^{-1} [\omega (\theta_j) - \omega_m (\theta_j)].
\end{equation}
Errors on the fit parameters are determined using $\Delta \chi^2 = 2.3$ and $\Delta \chi^2 =1$ for the two and one parameter fits, respectively.  We fit both over the full range that measurements are made, from 0.003$^{\circ}$ (slightly above the resolution of \textit{WISE}) to 2.1$^{\circ}$ (corresponding to $\sim$0.1- 60 Mpc at $z=1$), and for comparison over the same range as D13, 0.02$^{\circ}$-0.5$^{\circ}$  ($\sim$0.6 - 15 Mpc at $z=1$).  Fits over the full range are shown as dark dotted and dashed lines in Figues~\ref{fig:w4dex} and~\ref{fig:wbig}, while fits over the more limited range are shown as lighter colors.  All fit parameters are given in Table~\ref{table:fitsall}. Figure~\ref{fig:fitcompare} plots the fit parameters for easier visual comparison of the values in Table~\ref{table:fitsall}.

\begin{table*}
%  \scriptsize
  \caption{Fits to obscured and unobscured quasar autocorrelations.}
   \label{table:fitsall}
   \begin{tabular}{lccccccccccc}
    \hline
                                  &                            \multicolumn{5}{c}{Full Mask}  &  &  \multicolumn{5}{c}{Partial Mask}  \\
                                                         \cline{2-6}                                                                          \cline{8-12}                                                     \\
                                  &        \multicolumn{2}{c}{Full power-law}                 &  &    \multicolumn{2}{c}{Fixed-slope power-law} &  &  \multicolumn{2}{c}{Full power-law}                 &  &    \multicolumn{2}{c}{Fixed-slope power-law}         \\
                                                          \cline{2-3}                                                                                        \cline{5-6}            \cline{8-9}    \cline{11-12}                                                 \\  
Sample  & $A$      &     $\delta$       &   &        $A$        &       $\delta$  &   & $A$      &     $\delta$        &   &         $A$        &       $\delta$  \\
\hline
                                  &                                                  \multicolumn{5}{c}{0.003 - 2.1 degrees; 4/dex bins}  &  &               \multicolumn{5}{c}{0.003 - 2.1 degrees; 4/dex bins}                                               \\
                                                                                                           \cline{2-6}                                             \cline{8-12}                                                                              \\
Total                   & 0.0015 $\pm$ 0.0001  & $-$1.02 $\pm$ 0.01  &                 & 0.0016 $\pm$ 0.0000  &      $-1$   && 0.0013 $\pm$ 0.0001      & $-$1.18 $\pm$ 0.02 &            & 0.0019 $\pm$ 0.0001  &         $-1$                  \\
Obscured           & 0.0017 $\pm$ 0.0001  & $-$1.08 $\pm$ 0.02 &                 & 0.0022  $\pm$ 0.0001 &     $-1$  &&  0.0030 $\pm$ 0.0003       & $-$0.96 $\pm$ 0.06 &            & 0.0028 $\pm$ 0.0002 &          $-1$                  \\
Unobscured      & 0.0014 $\pm$ 0.0002  & $-$0.92 $\pm$ 0.04  &                & 0.0011 $\pm$ 0.0001  &     $-1$   &&  0.0015 $\pm$ 0.0002      & $-$0.95 $\pm$ 0.04  &           & 0.0013  $\pm$ 0.0001 &         $-1$                   \\
\hline  \\
                                  &                                                     \multicolumn{5}{c}{0.02 - 0.5 degrees; 4/dex bins}        &   &         \multicolumn{5}{c}{0.02 - 0.5 degrees; 4/dex bins}                                                              \\
                                                                                                               \cline{2-6}                                        \cline{8-12}                                                                                   \\
Total                    & 0.0020 $\pm$ 0.0002   & $-$0.94 $\pm$ 0.03   &              & 0.0017 $\pm$ 0.0001   &      $-1$   && 0.0013 $\pm$ 0.0001     & $-$1.24 $\pm$ 0.04     &           & 0.0023 $\pm$ 0.0001 &        $-1$              \\
Obscured           & 0.0027 $\pm$ 0.0004    & $-$0.92 $\pm$ 0.05   &             & 0.0022 $\pm$ 0.0002   &      $-1$   &&  0.0013 $\pm$ 0.0003     & $-$1.51 $\pm$ 0.09     &          & 0.0033 $\pm$ 0.0005  &        $-1$                \\
Unobscured      & 0.0017 $\pm$ 0.0004    & $-$0.92 $\pm$ 0.10   &             & 0.0014 $\pm$ 0.0002   &      $-1$  &&   0.0018 $\pm$ 0.0003     & $-$0.97 $\pm$ 0.07      &          & 0.0016 $\pm$ 0.0002 &         $-1$                \\
\hline  \\
                                  &                                                     \multicolumn{5}{c}{Two large bins}                             &   &           \multicolumn{5}{c}{Two large bins}                                                                       \\
                                                                                                               \cline{2-6}                                                \cline{8-12}                                                                               \\
Total                  & 0.0017 $\pm$ 0.0002     & $-$0.99 $\pm$ 0.04   &             & 0.0017 $\pm$ 0.0002   &      $-1$   &&   0.0018 $\pm$ 0.0002     & $-$1.17 $\pm$ 0.06     &          & 0.0023 $\pm$ 0.0002  &          $-1$          \\
Obscured         & 0.0026 $\pm$ 0.0005     & $-$0.95 $\pm$ 0.08   &             & 0.0024 $\pm$ 0.0003   &      $-1$   &&   0.0024 $\pm$ 0.0005      & $-$1.41 $\pm$ 0.11    &          & 0.0034 $\pm$ 0.0005   &         $-1$           \\
Unobscured    & 0.0011 $\pm$ 0.0002     & $-$1.08 $\pm$ 0.08   &             & 0.0013 $\pm$ 0.0002   &       $-1$  &&    0.0013 $\pm$ 0.0003     & $-$1.05 $\pm$ 0.07    &          & 0.0015 $\pm$ 0.0002   &         $-1$                 \\
\hline
     \end{tabular}
   \\  
{
\raggedright    
  The best fit power-laws (both with the slope fixed at $\delta = -1$ and with the amplitude and slope as free parameters).  The left side shows results utilizing the full mask, and the right half presents results without removing the regions around clustered points in the \textit{WISE} flagged data.  \\
 }
\end{table*}

\subsection{The Quasar Bias and Halo Masses}
Objects formed in the peaks of a Gaussian random field are expected to cluster more strongly than the underlying dark matter \citep[e.g.][]{1984ApJ...284L...9K, 1986ApJ...304...15B}.  This excess can in principle be dependent on scale, but is generally expected to be scale-independent in most scenarios on large scales.  This additional clustering signal is known as the quasar bias $b_q$, and relates the quasar autocorrelation $\omega_q$ to the dark matter autocorrelation $\omega_{dm}$: $\omega_q = \omega_{dm} b_q ^2$.

In order to calculate $\omega_{dm}$, we use the formulae of \citet{2003MNRAS.341.1311S} to first calculate the nonlinear, dimensionless power spectrum of dark matter ($\Delta^2(k,z)$, where $k$ is the wavenumber) using the $dN/dz$ described above \citep[see also the appendices of][]{2007ApJ...658...85M}.  The methods of \citet{2003MNRAS.341.1311S} give the dark matter power spectrum to an accuracy of $<3$\% at $z < 3$.  For small angles ($\theta \ll 1$ rad), we can project the power spectrum to an angular autocorrelation in a flat Universe using Limber's approximation \citep{1953ApJ...117..134L, Peebles:1980vn, 1991MNRAS.253P...1P}:
\begin{multline}
 \omega_{dm} (\theta) = \pi \int_{z=0}^{\infty} \int_{k=0}^{\infty} \frac{\Delta^2 (k,z)}{k} J_0 [k \theta \chi(z)] ~\times \\
    \left( \frac{dN}{dz} \right)^2 \left(\frac{dz}{d \chi} \right) \frac{dk}{k} dz.
\end{multline}
Here $J_0$ is the zeroth-order Bessel function of the first kind, $\chi$ is the comoving distance along the line of sight, $dN/dz$ is the normalized redshift distribution, and $dz/d \chi = H_z / c = (H_0 / c) [\Omega_m(1+z)^3 + \Omega_{\Lambda} ] ^{1/2}$ (valid for the flat cosmology used here).  We use a Monte-Carlo method to estimate this integral to a fractional accuracy of $<$1\%.  We calculate $\omega_{dm}$ three times, using the $dN/dz$ for the total, obscured and unobscured samples.  These are shown as the solid lines in Figures~\ref{fig:w4dex} and~\ref{fig:wbig}.

\begin{figure*}
\centering
\hspace{0cm}
   \includegraphics[width=14cm]{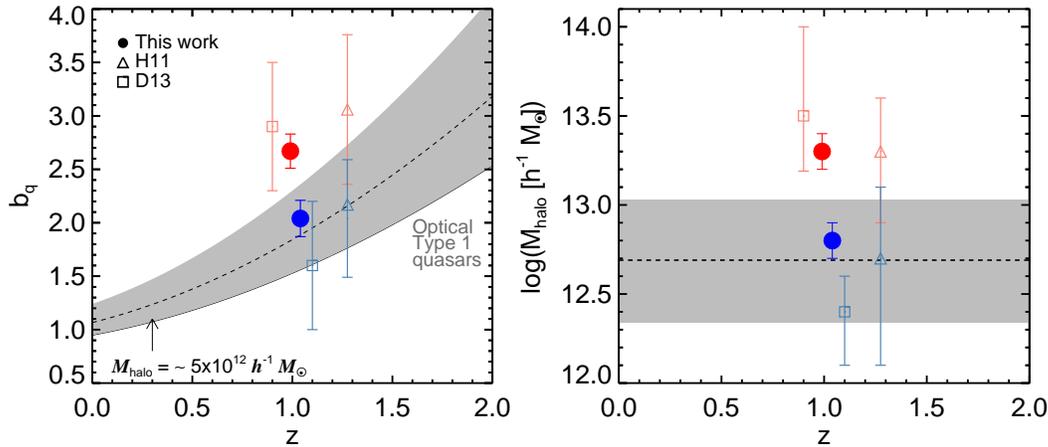}
   \vspace{0cm}
  \caption{\textit{Right:} A comparison of the bias values found in this study with a range of studies from the literature for unobscured quasars, as well as recent studies with split obscured/unobscured subsamples \citep[][D13]{2011ApJ...731..117H} . \textit{Left:} Same as the right panel, but comparing $M_{h}$.\label{fig:compare}}
\end{figure*}

To find $b_q$ we then rescale each $\omega_{dm}$ to fit $\omega_q = b_q^2 \omega_{dm}$ for the three samples, using the full covariance matrix in the fit as was done for the power-law fits.  For direct comparison with D13 we fit the bias over the range 0.04$^{\circ}$-0.4$^{\circ}$.  However, as we find with the full mask applied here the bias remains quite scale-independent to small angular scales, we can fit the bias over the larger range 0.01-1 degrees without changing the results significantly.  Errors on $b_q$ are determined from the fits where $\Delta \chi^2 =1$.  The quasar bias is shown as a function of scale in the lower panels of Figure~\ref{fig:w4dex} and~\ref{fig:wbig}, and listed in Table~\ref{table:bias}.  The bias values are shown in comparison with other values from the literature for unobscured sources in the right panel of Figure~\ref{fig:compare}.

It is clear in Figure~\ref{fig:compare} that the errors on $b_q$ here are smaller than those in D13, despite the fact that the error bars on $\omega_{\theta}$ are similar (though ours are smaller due to the broader binning, which could account for some of the difference in the $b_q$ errors), and the errors on $A$ and $\delta$ are consistent.  Some of the difference can be attributed the fact that our additional masking has improved the ability of a pure power-law to fit the data, especially if fits are performed only using the variance as opposed to the covariance.  We verify that the presented error bars are consistent with the variance in $b_q$ when fitting the results from each jackknife iteration.  Additionally, we shift all of the $\omega_{\theta}$ values up or down by their corresponding errors and refit the bias --- this is an overestimate of the bias errors, and only results in a shift of about 0.3.  Finally, the errors derived here are consistent with other angular autocorrelation measurements that use a similar jackknife technique, once differences in sample size and area are considered \citep[e.g.][]{2007ApJ...658...85M}.

Finally, we follow the method outlined in section 4.1 of \citet{2007ApJ...658...85M} to calculate dark matter halo masses ($M_h$) for each sample.  We refer the reader there for details, but this method uses the ellipsoidal collapse model of \citet{2001MNRAS.323....1S}.  Here, we model the linear power spectrum using the transfer function including the effect of baryons from \citet{1998ApJ...496..605E}.  We calculate the mass at the appropriate mean redshift for each sample, as listed in \S2.6.  The final values of the halo masses are listed in Table~\ref{table:bias}, and shown in comparison to literature values for unobscured quasars in the left panel of Figure~\ref{fig:compare}.

\begin{table*}
  \caption{Quasar bias and dark matter halo mass.}
  \label{table:bias}
  \begin{tabular}{lccccc}
  \hline
                                      &  \multicolumn{2}{c}{Full Mask}  &    &  \multicolumn{2}{c}{Partial Mask}    \\
                                                 \cline{2-3}                                                \cline{5-6}                                  \\
  Sample   &  $b_q$  &  $\log (M_{h}/M_{\odot}$ $h^{-1}$)  &   &  $b_q$  & $\log (M_{h}/M_{\odot}$ $h^{-1}$)   \\ 
\hline
				&	                          \multicolumn{5}{c}{4/dex bins}					\\
				                                                        \cline{2-6}                                  \\
   \vspace{0.1cm}
   Total                         & 2.39 $\pm$ 0.07 & $13.10^{+0.03}_{-0.04}$ & &  2.76 $\pm$ 0.09 & $13.31^{+0.05}_{-0.04}$            \\ 
   \vspace{0.1cm}
   Obscured                & 2.67 $\pm$ 0.16 & $13.29^{+0.10}_{-0.10}$ & &  3.02 $\pm$ 0.24 & $13.46^{+0.12}_{-0.10}$          \\
   \vspace{0.1cm}
   Unobscured           & 2.04 $\pm$ 0.17 & $12.84^{+0.14}_{-0.12}$ & &  2.22 $\pm$ 0.14 & $12.98^{+0.11}_{-0.08}$            \\                         
\hline \\
			        &	                          \multicolumn{5}{c}{Two large bins}					\\
				                                                        \cline{2-6}                                \\
 \vspace{0.1cm}
  Total                         & 2.43 $\pm$ 0.11 & $13.13^{+0.06}_{-0.06}$ & & 2.78 $\pm$ 0.13 & $13.32^{+0.09}_{-0.09}$           \\       
 \vspace{0.1cm}
   Obscured                & 2.84 $\pm$ 0.17 & $13.38^{+0.10}_{-0.07}$ & & 3.20 $\pm$ 0.23 & $13.54^{+0.08}_{-0.08}$              \\
 \vspace{0.1cm}
   Unobscured           & 1.96 $\pm$ 0.14 & $12.77^{+0.12}_{-0.10}$ & & 2.11 $\pm$ 0.13 & $12.90^{+0.08}_{-0.09}$             \\
\hline
   \end{tabular} 
    \\
{
\raggedright    
 The quasar bias, computed by fitting our model $\omega_{dm}$ (for each $dN/dz$) to $\omega_{\theta}$, and the corresponding dark matter halo masses ($M_h$) using the mean redshift of each subsample.  \\
 }
\end{table*}

\section{DISCUSSION \& CONCLUSIONS}
We find that with additional masking of the data, the difference between obscured and unobscured quasar angular clustering found in D13 is significantly reduced, for objects with a similar selection and in the same region of the sky.  We choose to bin the data slightly more heavily, preferring four bins per dex to the five bins per dex of D13 in order to reduce the errors.  The mask applied to the data by D13 falls roughly in between what we call here the ``full'' and ``partial'' masks.  This is because D13 remove regions around resolved \textit{2MASS} sources, which will not be removed using our partial mask, but will be incorporated as part of our full mask.  However, we note that the masks are different in other ways, for example in the way we handle regions of Moon contamination.

Given this, the slope of the best fit power-laws of our results for the \textit{total} quasar sample match the results of D13 fairly well.  We find a steeper slope using the partial mask, because of the highly clustered artifacts remaining and contributing to the signal on smaller scales.  For the fits where the slope is a free parameter, the results can change considerably depending on where (over which scales) the fits are performed, and there is degeneracy between the slope and amplitude (Figure~\ref{fig:fitcompare}).  Using our partial mask and fixing the slope at $\delta \sim -1$, we recover the D13 results for the clustering amplitude of obscured and unobscured sources --- $A_{obsc} \approx 0.003$ and $A_{unobsc} \approx 0.001$.  

It is clear however that with the partial mask, the assumption that $\delta \sim -1$ is not particularly valid.  The slope of the obscured quasar autocorrelation is markedly steeper than this, especially on smaller scales (Figures~\ref{fig:w4dex} and~\ref{fig:fitcompare}).  We also see, as in D13, that on small scales the quasar bias seems to increase with this mask (Figure~\ref{fig:w4dex}), more significantly in the obscured sample.   

The bias and halo mass values using the partial mask also agree well with the results of D13.  Although our unobscured bias is higher than theirs --- 2.0 compared to 1.6 --- it is within their error bar, and simply highlights that using independently developed masks can lead to different results.  Using the partial mask we find halo masses for obscured quasars roughly a factor of 5 larger than for unobscured quasars, as opposed to a factor of 10.

However, by applying the additional component of our mask that identifies regions around highly grouped points in the \textit{WISE} flagged data and removes them from the sample \textit{and} the random catalog, many of these issues are resolved.  The results are far less dependent on how the fits are performed (over which scales and with which free parameters).  The slope of the autocorrelation in all cases is roughly consistent with $\delta \sim -1$, and the scale dependence of the quasar bias is largely removed (Figure~\ref{fig:w4dex}).  The unobscured sample still retains $A \approx 0.001$, but the amplitude in the obscured sample is reduced to $A \approx 0.002$.  

When D13 limit their sample to only include objects with SDSS counterparts, they too find amplitudes in line with these values (their value of $A$ for the obscured sample decreases).  However, if we limit our sample to SDSS detected sources only, when using the full mask the results \textit{remain unchanged} (though the errors increase due to the decrease in sample size).  This strongly suggests that without the additional component in our mask, there are many \textit{WISE}-selected objects in the sample that are not actually AGN, but contaminants.  

Using our full mask for the \textit{WISE}-selected quasars produces values for the dark matter halo masses nearly identical to what is found in the \textit{Spitzer}-selected sample of \citet{2011ApJ...731..117H}.  We agree with their result that the halo mass of IR-selected unobscured quasars falls near the high end of results from optically selected samples, but is completely consistent (Figure~\ref{fig:compare}, right).  We find, as they do, that obscured quasars reside in halo masses $\sim$3 times as massive as those that host unobscured sources.  Due to our larger sample size, however, we confirm this difference at much higher significance ($\sim 4 \sigma$) and can say conclusively that obscured quasars cluster more strongly than unobscured sources, and thus reside in higher-mass halos.

As mentioned in \S 2.6, it is possible that the spike in the redshift distribution of the unobscured quasars is problematic.  These objects may be resolved sources which have a significant contribution from starlight in the nuclear regions, leading to their classification as unobscured when in fact they are obscured \citep[e.g.][]{2007ApJ...671.1365H}.  At high redshift, where sources are unlikely to be resolved, this is not likely to be an issue as the starlight should be too faint.  Therefore, to test the effects of this possibility, we repeat our analysis using only point-like sources in the SDSS imaging of the unobscured sample.  We also re-calculate $\omega_{dm}$ with this spike in $dN/dz$ removed, and re-measure the bias.  We find that this decreases the mean redshift in the unobscured sample to $\sim 1.2$, and increases the bias slightly to $b_q \sim 2.2$.  If the sources removed using this method should in fact be part of the obscured sample, then we can assume that their average redshift and bias will \textit{decrease} in a similar way, to $\sim 2.6$, reducing the measured difference in bias reported here.  However, this has a minimal effect on the resulting difference in halo masses because of the changes in mean redshift.  

While these measurements improve the absolute determinations of the IR-selected obscured and unobscured quasar bias and halo mass, the conclusions are the same as previous work in this area.  These results essentially rule out the simplest, orientation-only (``dusty-torus'') models for the obscured quasar population, though it is certainly possible that orientation plays a role and is responsible for some fraction of the population.  It is also possible that the obscuring material is found in large (galaxy-scale) structures and the covering fraction and structure is driven by environment.  It is most likely that material both in the nuclear region and at larger scales is responsible for obscuration in the population as a whole.

A difference in clustering strength for obscured and unobscured quasars may also suggest an evolutionary effect (or again, some combination of orientation and evolution).  If a picture such as that from \citet{2008ApJS..175..356H} is correct, where obscured sources are found in a ``pre-blowout'' phase, then obscured quasars would be a younger version of unobscured sources.  This interpretation requires us to assume that both the range  of bolometric luminosities in our samples are the same, which is not necessarily unreasonable given the similarity in $W2$ flux and redshift, and the Eddington rates are similar.  These combined would indicate that the ranges in black hole masses are the same, and thus correlations between black hole mass and halo mass \citep[e.g.][]{2002ApJ...578...90F, 2010MNRAS.405L...1B, 2011Natur.469..377K} would suggest that they should have the same halo mass --- if the black holes are at their final mass.  The stronger clustering/larger halo masses would then be more evidence that obscured quasars are young and in the process of ``catching up'' in black hole mass.  This would thus imply that black hole growth lags behind that of the halo, a question of active study \citep[e.g.][]{2006ApJ...649..616P, 2008AJ....135.1968A, 2008ApJ...681..925W, 2010MNRAS.402.2453D, 2013ARA&A..51..511K}.

Abundance matching techniques \citep[e.g.][]{1999ApJ...523...32C, 1999ApJ...520..437K, 2004MNRAS.353..189V, 2006ApJ...643...14S, 2010MNRAS.404.1111G} which essentially compare the number density of a quasar population to the number density of halos at the typical parent halo mass, can constrain the length of the active quasar phase.  Using the bolometric quasar luminosity function of \citet{2007ApJ...654..731H} at $z\sim1$ and the median luminosity of typical IR-selected quasars at this flux limit of $L_{\rm bol}\sim10^{46}$ erg s$^{-1}$ \citep[e.g.][Hainline et al.\, in prep]{2011ApJ...731..117H}  , we obtain a space density of WISE-selected quasars of $\sim$$2\times10^{-5}$ Mpc$^{-1}$, of which $\approx$60\% are unobscured and $\approx$40\% obscured as determined in \S2.4. For the best-fit dark matter halo masses obtained above, the halo space densities are $dn/d\log_{10}(M)=(4 \pm 1)\times10^{-4}$ Mpc$^{-3}$ and $(1\pm0.5)\times10^{-4}$ Mpc$^{-3}$, respectively. The bulk of our sample lies in the redshift range $0.5 < z < 1.5$ that spans $\sim$4 Gyr of cosmic time, so from the resulting occupation fractions we obtain lifetimes of $\sim 140\pm40$ Myr for unobscured quasars and $\sim 320\pm120$ Myr for obscured quasars.

Both of these lifetime estimates are consistent with previous results from the clustering of unobscured objects ($\sim 10^7$ to $10^8$ years), but in combination the results are in 2$\sigma$ tension with models where the obscured and unobscured phases are the same length, suggesting instead that the obscured phase is of the order a few times longer than the unobscured phase.  We still find that the lifetime of either phase is of order a few percent of the Hubble time.

Deciphering how large of a role orientation, evolutionary state, and quasar lifetime play in the differences between obscured and unobscured quasars requires accurate determination of their bias and halo mass --- this work provides an improved mask for \textit{WISE} data toward this end.  Future work on the modeling of the observed clustering of the obscured quasar population should focus on the results presented here or in \citet{2011ApJ...731..117H}, as the stronger clustering amplitudes found in other analyses of \textit{WISE}-selected AGN are likely a systematic effect due to insufficient masking of the data.  However, all of these results may suffer from systematics because of the dependence on the angular mask.  An even more ideal way to measure the bias of \textit{WISE}-selected obscured and unobscured quasars, and hopefully confirm these results, would be to use the mask-independent method of cross-correlating the quasar density with the lensing convergence of the cosmic microwave background \citep[e.g.][]{2012PhRvD..86h3006S, 2013ApJ...776L..41G}.  We will present such an analysis in a future paper.

\section*{Acknowledgements}
MAD, ADM, RCH and KNH were partially supported by NASA through ADAP award NNX12AE38G and EPSCoR award NNX11AM18A and by the National Science Foundation through grant numbers 1211096 and 1211112. JEG acknowledges support from the Royal Society.

\bibliography{full_library.bib}

\label{lastpage}

\end{document}

%% file: commands_mnras.tex
\newcommand\ion[2]{#1$\;${\small\rmfamily\@Roman{#2}}\relax}%
\def\lsim{\lower0.3em\hbox{$\,\buildrel <\over\sim\,$}}
\def\gsim{\lower0.3em\hbox{$\,\buildrel >\over\sim\,$}}